\documentclass[aps,prd,11pt,notitlepage,nofootinbib,superscriptaddress,showkeys,showpacs]{revtex4-2}
\pdfoutput=1
\usepackage{amstext,amsmath,amssymb,amsfonts,bbm,euscript,color,dsfont}
\usepackage[english]{babel}
\usepackage[utf8]{inputenc}
\usepackage{epsfig}
\usepackage{hyperref}
\usepackage{amsthm}
\usepackage{subfigure}
\usepackage{color}
\usepackage{multirow}
\usepackage{bbold}
\usepackage{tikz}
\usetikzlibrary{arrows,backgrounds}
\usepackage{verbatim} 
\usepackage{graphicx}
\usepackage{latexsym}

\makeatother

\begin{document}

\title{{\bf\Large The supersymmetric extension of the replica model}}

\author{M.~A.~L.~Capri}\email{caprimarcio@gmail.com}
\affiliation{Universidade do Estado do Rio de Janeiro (UERJ),
Instituto de F\'{i}sica, Departamento de F\'{i}sica Te\'{o}rica,
Rua S\~ao Francisco Xavier 524, Maracan\~{a}, Rio de Janeiro, Brasil, CEP 20550-013}

\author{R.~C.~Terin} 
\email{rodrigorct@ita.br}
\affiliation{Instituto Tecnol\'{o}gico de Aeron\'{a}utica (ITA), DCTA, Departamento de F\'{i}sica, Pra\c{c}a Marechal Eduardo Gomes, 50 - Vila das Ac\'{a}cias, S\~{a}o Jos\'{e} dos Campos, S\~{a}o Paulo, Brasil, CEP 12228-900}

\author{H.~C.~Toledo}
\email{henriqcouto@gmail.com}
\thanks{Dr. H. C. Toledo was affiliated to UERJ, as a PhD student, at the time of the development of this work.}
\affiliation{Universidade do Estado do Rio de Janeiro (UERJ),
Instituto de F\'{i}sica, Departamento de F\'{i}sica Te\'{o}rica,
Rua S\~ao Francisco Xavier 524, Maracan\~{a}, Rio de Janeiro, Brasil, CEP 20550-013}

\begin{abstract}
\noindent We perform a $\mathcal{N}=1$ supersymmetric extension of the replica model quantized in the Landau gauge and compute the gluon and gluino propagators at tree-level, such results display a supersymmetric confined model very similar to the supersymmetric version of the Gribov-Zwanziger approach.  
\end{abstract}
	
\maketitle
\section{Introduction}

The well-known Gribov problem \cite{Gribov:1977wm} has been under research in past decades for attempting to explain in a consistent way the quantization of nonabelian gauge theories. Nowadays, one can understand that such issue is somehow connected with confinement's phenomenon of quarks and gluons. From this background picture, some models have been proposed in order to improve the so-called Faddeev-Popov's (FP) mechanism \cite{Faddeev:1967fc}. One of them is the (refined) Gribov-Zwanziger (RGZ) \cite{Zwanziger:1989mf,Zwanziger:1990tn,Dudal:2008sp} approach whose construction is based on the fact that  the integral over gauge field configurations is restricted to the first Gribov region, where the FP operator is positive definite. However, also other first principles approaches were proposed during this century like \cite{sorella2011,Serreau:2012cg,Serreau:2015yna,Reinosa:2020skx} and, in this context, the replica model appears as an interesting alternative. In fact, as showed in \cite{Capri:2010pg}, a nice result obtained from the replica model is the construction of composite operators of the lightest glueball states in pure  $SU(N)$ Yang-Mills (YM) theory in Landau gauge. 

Such proposal emerged on an attempt to study confining aspects of YM theory without introducing the so-called Gribov regions. In this model, the gauge field interacts with its replica. This interaction softly breaks the well-known Becchi-Rouet-Stora-Tyutin (BRST) symmetry generating a structure for gluon's propagator that does not obey the Källén-Lehmann spectral representation \cite{Lehmann:1954xi} \footnote{This construction follows the original idea of confined models having a BRST symmetry soft breaking, however, currently there is a new formulation based on the Stückelberg field which is responsible for the BRST-invariance of the extras terms that emerges in the infrared regime of YM theory, for instance see \cite{Capri:2019drm} and references quoted therein.}. Moreover, the replica model does not exclude the presence of the Gribov copies, whereas the appearance of those extra gauge field configurations is directly related to the null eigenvalues of the FP operator\footnote{For more details on the presence of Gribov copies for such proposal read \cite{sorella2011}.}.  

Since gluon's propagator behavior is the core of all discussions of confinement's phenomenon, the use of new insights obtained from lattice simulations are responsible for feed up the first principle models. For instance, the massive behavior for the gluon propagator in deep infrared energy sector was an outstanding result achieved by lattice, see \cite{Cucchieri:2004mf,Cucchieri:2007md,Cucchieri:2007rg,Cucchieri:2008mv} and references quoted therein. Indeed,  by taking  into account the restriction to the first Gribov region, the gluon propagator at tree level is modified in a way that its result 
\begin{equation}
\langle A^{a}_{\mu}(k)A^{b}_{\nu}(-k)\rangle=\delta^{ab}\frac{k^{2}}{k^{4}+\gamma^{4}}P_{\mu\nu}(k)\,,
\label{Ggribov}
\end{equation} 
gives the positivity violation of gluon's spectral density \cite{Cornwall:2013zra}, such result is one of the arguments nowadays to define gluon as a nonphysical excitation. Moreover, in Eq.~\eqref{Ggribov}, one can see that the internal symmetry group $SU(N)$ is employed in its adjoint representation with latin indices ($a,b,c, \dots$) running over $1$ to ($N^{2}-1$). Also, one has the massive Gribov parameter $\gamma$ and the transverse projector $P_{\mu\nu}$. Furthermore, one can easily note that the Gribov-type propagator \eqref{Ggribov} has complex poles, namely, $\pm i\gamma^{2}$. In fact, it can be decomposed in terms of two nonphysical modes
with complex conjugate imaginary masses known as $i$-particles. Such property, which is also present in the replica model as we shall see later, indicates the nonphysical character of the gluon as a genuine physical particle, but, at the same time, provides us to construct
examples of local composite operators whose correlation functions have good analyticity properties, as expressed by the K\"all\'{e}n-Lehmann spectral representation, being identified with the glueball states.


On the other hand, the pure $\mathcal{N}=1$ super Yang-Mills theory has some correspondence with Quantum Chromodynamics (QCD) with one flavor of quarks \footnote{The difference here remains on the fact that the gluino is in the adjoint representation of the gauge symmetry group.}, indeed, in order to shed some light on the confinement phenomenon, there are some interesting works in the literature like \cite{VENEZIANO1982231,NOVIKOV1986329,SHIFMAN1988445}, where the issues of a low effective action responsible for describing the dynamics of composite operators, the exact beta function and the gluino condensate were addressed. Moreover, a subject matter under investigation nowadays is determine the role played by the Gribov problem on the quantization of supersymmetric nonabelian models. To accomplish such goal, some studies have been done since last decade, see \cite{Amaral:2013uya,Capri:2014xea} and references therein.


Therefore it would be interesting to address the Gribov problem in gauge-fixing quantization for supersymmetric nonabelian gauge models by taking a look only at the effects at nonperturbative sector, whose characteristic to be exploited in such case is the confinement phenomenon of the elementary degrees of freedom (namely, gluons and gluinos) by computing their two-point correlation functions. Thus, the supersymmetric extension of the replica model, or simply supersymmetric replica model (SRM), is one of the proposals to fulfill such aim since there are some lattice simulations studies following this direction, {\it e.g.} \cite{Farchioni:2004fy} and references therein, which offers us a possibility to compare different viewpoints\footnote{ The discrete one from the lattice and the continuum one from our approach.} in the near future.  

Also in the near future, we would like to provide with the SRM a distinct perspective from other approaches in continuum (like supersymmetric GZ) with respect of possible non-perturbative features as the gluino condensate emergence, glueball's spectrum, and a possible understanding about the inclusion of fermions in non-supersymmetric replica model, which remains as an open problem until now. Therefore, in this article our aim is twofold: $(i)$ We will extend the replica model to its supersymmetric version, where the vector superfield $V$ interacts with its own replica, namely $\hat{V}$ due to a massive parameter (responsible for coupling both superfields), generalizing the non-supersymmetric version given in  \cite{sorella2011}; $(ii)$ As a consequence of task number one, we can verify, by direct comparison, if the superprogators obtained here are in qualitative agreement with propagators' results achieved from super Gribov-Zwanziger's action established in \cite{Amaral:2013uya}. 

The paper is organized as follows. After this brief introduction, in Section \ref{Rev}, we review the non-supersymmetric version of the replica model. In Section \ref{susy}, we propose a $\mathcal{N}=1$ supersymmetric version for such framework, the so-called SRM. In Section \ref{prop}, we compute the gluon and gluino propagators at tree-level. Finally, in Section \ref{concl}, we display our main results, conclusions and perspectives. For the benefit of the reader, we also display in Appendix \ref{conv} 
the notations and conventions used in this manuscript.
\section {The replica model review}
\label{Rev}

We will present here in this Section a brief review of the original (non-supersymmetric) replica model. Such approach could be considered as a different viewpoint to interpret correctly the confined states when compared with other first principles frameworks whose construction is based on Gribov copies \cite{Gribov:1977wm, Zwanziger:1989mf, Serreau:2012cg, Reinosa:2020skx}. The starting point is the well-known pure YM action quantized by the FP mechanism in Euclidean space, i.e. 
\begin{eqnarray}
\mathcal{S}_{YMFP} &=& \int d^{4}x\,\bigg(\frac{1}{4g^{2}}\,F^{a}_{\mu\nu}F^{a}_{\mu\nu}
+ib^{a}\,\partial_{\mu}A^{a}_{\mu}
+\bar{\eta}^{a}\,\partial_{\mu}D^{ab}_{\mu}\eta^{b}\bigg)\,.
\label{pureYangmills}
\end{eqnarray}
Hereby, we have adopted the latin letters $\{a,b,c,\dots\}$, running from 1 to $N^{2}-1$, representing the SU($N$) group. Moreover, the Nakanishi-Lautrup field $b^{a}$ plays a role of a Lagrange multiplier, therefore it defines in a off-shell fashion the Landau gauge-fixing condition, $\partial_{\mu}A^{a}_{\mu}=0$. Also we have the Faddeev-Popov ghosts, $(\bar{\eta}^{a},\eta^{a})$, and finally the YM field strength $F^{a}_{\mu\nu}$ and the covariant derivative $D_{\mu}^{ab}$  whose definitions for this work are
\begin{equation}
F_{\mu\nu}^{a}=\partial_{\mu}A^{a}_{\nu}-\partial_{\nu}A^{a}_{\mu} + f^{abc}A_{\mu}^{b}A_{\nu}^{c}\,,
\end{equation}
\begin{equation}
D_{\mu}^{ab}=\partial_{\mu}\delta^{ab}-f^{abc}A_{\mu}^{c}\,.
\end{equation}
Additionally, as follows, the action \eqref{pureYangmills} is left invariant by a set of BRST transformations 
\begin{eqnarray}
sA_{\mu}^{a} &=& -D_{\mu}^{ab}\eta^{b}\,,\nonumber\\
s\eta^{a}&=&\frac{1}{2}f^{abc}\eta^{b}\eta^{c}\,,\nonumber\\
s\bar{\eta}^{a} &=& ib^{a}\,,\nonumber\\
sb^{a} &=& 0\,,
\label{BRST}
\end{eqnarray}
where $s$ is the nilpotent $(s^{2}=0)$ BRST variation operator. As usual, in order to deal with the nonlinear BRST transformations of the gauge $A_{\mu}^{a}$ and the ghost $\eta^{a}$ fields one needs to introduce in the action \eqref{pureYangmills} a set of external sources $(\Omega^{a},L^{a})$, therefore we define the following term to be added to action \eqref{pureYangmills}:
\begin{eqnarray}
\mathcal{S}_{sources} &=& s\,\int d^{4}x \bigg(\Omega^{a}_{\mu}A^{a}_{\mu} + L^{a}\eta^{a}\bigg)\nonumber\\
&=& \int d^{4}x\,\bigg(
-\Omega^{a}_{\mu}\,D^{ab}_{\mu}\eta^{b}
+\frac{1}{2}f^{abc}L^{a}\eta^{b}\eta^{c}\bigg)\,,
\end{eqnarray}
such that these external sources transform as BRST singlets,
\begin{equation}
s\Omega_{\mu}^{a} = sL^{a} = 0\,.
\end{equation}
Thus, we define the most general action for pure YM theory, in Euclidean space, quantized in the Landau gauge as
\begin{eqnarray}
\mathcal{S} &=& \mathcal{S}_{YMFP}+\mathcal{S}_{sources}\nonumber\\ \mathcal{S} &=&\int d^{4}x\,\bigg(\frac{1}{4g^{2}}\,F^{a}_{\mu\nu}F^{a}_{\mu\nu}
+ib^{a}\,\partial_{\mu}A^{a}_{\mu}
+\bar{\eta}^{a}\,\partial_{\mu}D^{ab}_{\mu}\eta^{b}-\Omega^{a}_{\mu}\,D^{ab}_{\mu}\eta^{b}
+\frac{1}{2}f^{abc}L^{a}\eta^{b}\eta^{c}\bigg)\,.
\label{pureYangmillscomplete}
\end{eqnarray}
This action is suitable to the algebraic analysis of the renormalizability of the theory by means of the Slavnov-Taylor and Ward identities \cite{Piguet:1995er}.

At this point, we are almost ready to present the replica model, but let us consider first a very similar theory to the original YMFP, given by action Eq.~\eqref{pureYangmillscomplete}\footnote{The mirror field strength $\hat{F}^{a}_{\mu\nu} = \partial_{\mu}\hat{A}^{a}_{\nu}-\partial_{\nu}\hat{A}^{a}_{\mu} + f^{abc}\hat{A}_{\mu}\hat{A}_{\nu}$ brings us the existence of only one coupling constant. Such conclusion is made since the existence of the mirror symmetry invokes $ g\equiv \hat{g}$, thus it is possible to transform one gauge field into another.},
\begin{eqnarray}
\hat{\mathcal{S}} = \int d^{4}x\,\bigg(\frac{1}{4g^{2}}\,\hat{F}^{a}_{\mu\nu}\hat{F}^{a}_{\mu\nu}
+i\hat{b}^{a}\,\partial_{\mu}\hat{A}^{a}_{\mu}
+\hat{\bar{\eta}}^{a}\,\partial_{\mu}\hat{D}^{ab}_{\mu}\hat{\eta}^{b}
-\hat{\Omega}^{a}_{\mu}\,\hat{D}^{ab}_{\mu}\hat{\eta}^{b}
+\frac{1}{2}f^{abc}\hat{L}^{a}\hat{\eta}^{b}\hat{\eta}^{c}\bigg)\,.
\label{pureYangmillsR}
\end{eqnarray}
In fact, the actions \eqref{pureYangmillscomplete} and \eqref{pureYangmillsR} are exactly the same, except for the ``hat'' $(\,\hat{\/}\,)$ notation adopted for fields and sources of action \eqref{pureYangmillsR}. Then one can say that action \eqref{pureYangmillsR} is a {\it replica} of \eqref{pureYangmillscomplete}. Besides, action \eqref{pureYangmillsR} is left invariant by its own set of BRST transformations:
\begin{eqnarray}
\hat{s}\hat{A}_{\mu}^{a} &=& -\hat{ D}_{\mu}^{ab}\hat{\eta}^{b}\,=\, -(\partial_{\mu}\delta^{ab} - f^{abc}\hat{A}^{c}_{\mu})\hat{\eta}^{b} \,,\nonumber\\
\hat{s}\hat{\eta}^{a}&=&\frac{1}{2}f^{abc}\hat{\eta}^{b}\hat{\eta}^{c}\,,\nonumber\\
\hat{s}\hat{\bar{\eta}}^{a} &=& i\hat{b}^{a}\,,\nonumber\\
\hat{s}\hat{b}^{a} &=& 0\,,\nonumber\\
\hat{s}\hat{\Omega}^{a}_{\mu}&=&0\,,\nonumber\\
\hat{s}\hat{L}^{a}&=&0\,.
\label{BRST_hat}
\end{eqnarray}
Both theories are softly coupled one each other through a massive-like parameter, i.e.
\begin{equation}
\mathcal{S}_{\vartheta} = i\vartheta\sqrt{2}\int d^{4}x \bigg(A^{a}_{\mu}\hat{A}^{a}_{\mu}\bigg)\,,
\label{vartheta}
\end{equation}
where $\vartheta$ is a massive parameter,  analogous to the Gribov parameter $\gamma$ in the GZ framework \cite{Gribov:1977wm,ZWANZIGER1989591,Zwanziger:1989mf}, as will be clear later, and the full {\it replica model} is then given by
\begin{eqnarray}
\mathcal{S}_{full} &=& \mathcal{S} + \hat{\mathcal{S}} + \mathcal{S}_{\vartheta}\nonumber\\
&=& \int d^{4}x\,\bigg(\frac{1}{4g^{2}}\,F^{a}_{\mu\nu}F^{a}_{\mu\nu}
+ib^{a}\,\partial_{\mu}A^{a}_{\mu}
+\bar{\eta}^{a}\,\partial_{\mu}D^{ab}_{\mu}\eta^{b}-\Omega^{a}_{\mu}\,D^{ab}_{\mu}\eta^{b}
+\frac{1}{2}f^{abc}L^{a}\eta^{b}\eta^{c}\bigg)\nonumber\\
&+&\int d^{4}x\,\bigg(\frac{1}{4g^{2}}\,\hat{F}^{a}_{\mu\nu}\hat{F}^{a}_{\mu\nu}
+i\hat{b}^{a}\,\partial_{\mu}\hat{A}^{a}_{\mu}
+\hat{\bar{\eta}}^{a}\,\partial_{\mu}\hat{D}^{ab}_{\mu}\hat{\eta}^{b}
-\hat{\Omega}^{a}_{\mu}\,\hat{D}^{ab}_{\mu}\hat{\eta}^{b}
+\frac{1}{2}f^{abc}\hat{L}^{a}\hat{\eta}^{b}\hat{\eta}^{c}\bigg)\nonumber\\
&+&i\vartheta\sqrt{2}\int d^{4}x \bigg(A^{a}_{\mu}\hat{A}^{a}_{\mu}\bigg)\,.
\label{fullaction}
\end{eqnarray}
Action \eqref{fullaction} is invariant by  the so-called {\it mirror symmetry} whose definition is
\begin{eqnarray}
\Big(A^{a}_{\mu},b^{a},\eta^{a},\bar{\eta}^{a},L^{a},\Omega^{a}_{\mu}\Big) &\longrightarrow& \Big(\hat{A}^{a}_{\mu},\hat{b}^{a},\hat{\eta}^{a},\hat{\bar{\eta}}^{a},\hat{L}^{a},\hat{\Omega}^{a}_{\mu}\Big)\,,\nonumber\\
\Big(\hat{A}^{a}_{\mu},\hat{b}^{a},\hat{\eta}^{a},\hat{\bar{\eta}}^{a},\hat{L}^{a},\hat{\Omega}^{a}_{\mu}\Big)
&\longrightarrow&
\Big(A^{a}_{\mu},b^{a},\eta^{a},\bar{\eta}^{a},L^{a},\Omega^{a}_{\mu}\Big)\,.
\end{eqnarray} 
Then, the results for the tree-level propagators obtained from action \eqref{fullaction} are written as \cite{sorella2011}, 
\begin{eqnarray}
\langle A^{a}_{\mu}(k)A_{\nu}^{b}(-k)\rangle &=& \delta^{ab}\frac{k^{2}}{k^{4} + 2\vartheta^{4}}P_{\mu\nu}(k)\,,\nonumber\\
\langle \hat{A}^{a}_{\mu}(k)\hat{A}_{\nu}^{b}(-k)\rangle &=& \delta^{ab}\frac{k^{2}}{k^{4} + 2\vartheta^{4}}P_{\mu\nu}(k)\,,\nonumber\\
\langle A^{a}_{\mu}(k)\hat{A}_{\nu}^{b}(-k)\rangle &=& \delta^{ab}\bigg(\frac{-i\sqrt{2}\vartheta^{2}}{k^{4} + 2\vartheta^{4}}\bigg)P_{\mu\nu}(k)\,.
\end{eqnarray}
As one can immediately see, these propagators have a GZ-type structure. Therefore, the fields $A_{\mu}$ and $\hat{A}_{\mu}$ do not have an interpretation of physical particle excitations. Furthermore, one may note that at deep ultraviolet limit i.e., $\vartheta = 0$, the actions \eqref{pureYangmillscomplete} and \eqref{pureYangmillsR} decouple from each other, remaining two independent YMFP actions. This is analogous to a system of two identical harmonic oscillators coupled to a spring. The mass parameter $\vartheta$ is similar to the spring constant, usually denoted by $k$.  Then the ultraviolet limit $\vartheta\rightarrow 0$ is equivalent to a weak coupling $k\rightarrow0$.

The presence of the coupling term \eqref{vartheta} softly breaks both BRST symmetries enjoyed by actions \eqref{pureYangmillscomplete} and \eqref{pureYangmillsR}. Indeed, we have,
\begin{eqnarray}
s\mathcal{S}_{\vartheta}&=&-i\vartheta\sqrt{2}\int d^{4}x\,(D^{ab}_{\mu}\eta^{b})\hat{A}^{a}_{\mu}\,,\nonumber\\
\hat{s}\mathcal{S}_{\vartheta}&=&-i\vartheta\sqrt{2}\int d^{4}x\,A^{a}_{\mu}(\hat{D}^{ab}_{\mu}\hat{\eta}^{b})\,,
\end{eqnarray}
where we have assumed that
\begin{eqnarray}
\hat{s}\Big(A^{a}_{\mu}, b^{a},\eta^{a},\bar{\eta}^{a},\Omega^{a}_{\mu},L^{a}\Big)&=&0\,,\nonumber\\
{s}\Big(\hat{A}^{a}_{\mu}, \hat{b}^{a},\hat{\eta}^{a},\hat{\bar{\eta}}^{a},\hat{\Omega}^{a}_{\mu},\hat{L}^{a}\Big)&=&0\,.
\end{eqnarray}
However, being soft\footnote{By {\it soft} we mean the breaking has dimension less than the space, in this case the four dimensional Euclidean space.}, the breaking term can be conveniently rewritten into a BRST invariant fashion by means of the introduction of extra external sources and then the renormalizability of action \eqref{fullaction} can be established as already done in \cite{Capri:2010pg}.

Now we are ready to extend the replica model reviewed here to a supersymmetric formulation in $\mathcal{N}=1$ superspace.

\section{The supersymmetric extension of the replica model}
\label{susy}
In this section, we are going to present the supersymmetric version of the replica model in $\mathcal{N} = 1$ superspace, i.e. the  so-called SRM. As a first step, we ought to set up our conventions by first reviewing the super Yang-Mills (SYM) action quantized in Landau's gauge. 

\subsection{The $\mathcal{N} = 1$ Euclidean SYM action on superspace in the Landau gauge}

Let us start here a brief review of the pure $\mathcal{N}=1$ Euclidean SYM action on superspace. Its classical formulation, i.e. before the FP gauge-fixing procedure, is given by
\begin{equation}
\mathcal{S}_{SYM}=\mathcal{S}+\bar{\mathcal{S}}\,,
\end{equation}
with
\begin{equation}
\mathcal{S}=\frac{1}{128g^{2}}\,Tr\int d^{4}x\,d^{2}\theta\,W^{\alpha}W_{\alpha}
\end{equation}
and
\begin{equation}
W_{\alpha}=\bar{\mathcal{D}}^{2}\left(e^{-gV}\mathcal{D}_{\alpha}e^{gV}\right)\,.
\end{equation}
Also, $\bar{\mathcal{S}}$ is the Osterwalder-Schrader (OS) conjugate of $\mathcal{S}$ \cite{Lukierski:1982hr,Osterwalder:1972vwp,Osterwalder:1973zr,Frohlich:1974zs,Zumino:1977yh,Nicolai:1978vc,Kupsch:1988sb,Wetterich:2010ni,Amaral:2013uya}, and  $\mathcal{D}_{\alpha}$ and $\bar{\mathcal{D}}_{\dot{\alpha}}$ (with $\alpha=1,2$ and $\dot{\alpha}=\dot{1},\dot{2}$) are known as the chiral and antichiral covariant derivatives written as
\begin{eqnarray}
\mathcal{D}_{\alpha} &=& \frac{\partial}{\partial\theta^{\alpha}} + i\sigma^{\mu}_{\alpha \dot{\alpha}}\bar{\theta}^{\dot{\alpha}}\partial_{\mu}\,,\\
\bar{\mathcal{D}}_{\dot{\alpha}} &=& -\frac{\partial}{\partial\bar{\theta}^{\dot{\alpha}}} - i\theta^{\alpha}\sigma^{\mu}_{\alpha \dot{\alpha}}\partial_{\mu}\,.
\end{eqnarray}
Finally, from the eqs. above we have also the presence of the vector superfield $V(x,\theta,\bar{\theta})$, which may be described in SU($N$) group with reference to the group generators as
\begin{equation}
V(x,\theta,\bar{\theta}) = \sum_{a=1}^{N^{2}-1}V^{a}(x,\theta,\bar{\theta})T^{a}\,,\\
\end{equation}
having $T^{a}$ fixed by $N^{2}-1$ group generators of SU($N$), $x$ the four dimensional Euclidean space coordinate and $(\theta,\bar{\theta})$ the superspace coordinates. In addition, the general expression for the vector superfield in its components is given by
\begin{eqnarray}
V^{a} & = & C^{a}(x)+\theta^{\alpha}\chi^{a}_{\alpha}(x)+\bar{\theta}_{\dot{\alpha}}\bar{\chi}^{a\dot{\alpha}}(x)+\frac{1}{2}\theta^{2}M^{a}(x)+\frac{1}{2}\bar{\theta}^{2}\bar{M}^{a}(x)+\theta^{\alpha}\sigma_{\alpha\dot{\alpha}}^{\mu}\bar{\theta}^{\dot{\alpha}}A^{a}_{\mu}(x)\nonumber\\
&+&\frac{1}{2}\bar{\theta}^{2}\theta^{\alpha}\lambda^{a}_{\alpha}(x)+\frac{1}{2}\theta^{2}\bar{\theta}_{\dot{\alpha}}\bar{\lambda}^{a\dot{\alpha}}(x)+\frac{1}{4}\theta^{2}\bar{\theta}^{2}\mathfrak{D}^{a}(x)\,.
\label{V}
\end{eqnarray}
Hereby, the set $\{C,\chi_{{\alpha}},\bar\chi_{\dot{\alpha}},M,\bar{M},A_{\mu},\lambda_{\alpha},\bar\lambda_{\dot\alpha},\mathfrak{D}\}$ represents the superfield components in the adjoint representation of SU($N$) group\footnote{We also have in Eq.~\eqref{V} that $\theta^{2}=\theta^{\alpha}\theta_{\alpha}$, $\bar\theta^{2}=\bar{\theta}_{\dot{\alpha}}\bar{\theta}^{\dot{\alpha}}$ and $\sigma^{\mu}=(\mathbf{1},\sigma_1,\sigma_2,\sigma_3)$, being $\mathbf{1}$ the identity matrix of order two and $(\sigma_1,\sigma_2,\sigma_3)$ the usual Pauli matrices.}$^{,}$\footnote{The notation and convention were extracted from \cite{Amaral:2013uya} and \cite{Piguet:1996ys}. We also displayed an appendix at the end of this paper in order to clarify such subject. The Euclidean formulation is also a particular point we have taken into account very carefully.}. Moreover, fields $A_{\mu}^{a}$ and $(\lambda_{\alpha},\bar\lambda_{\dot\alpha})$ are known as the gluon and gluino fields, respectively \footnote{The gluino is its own antiparticle since the fields $\lambda_{\alpha}$ and $\bar\lambda_{\dot{\alpha}}$ are Majorana fermions.}.  Nonetheless, in order to quantize a supersymmetric gauge field theory one needs to implement an extra constraint, namely the gauge-fixing condition   
\begin{eqnarray}
\bar{\mathcal{D}}^{2}\mathcal{D}^{2}V^{a} &=& 0 \,.
\label{SUPERFPYM1}
\end{eqnarray} 
Also, from the FP quantization method for supersymmetry the Landau gauge-fixing condition is given by
\begin{eqnarray}
\mathcal{S}_{gf} & = & \frac{1}{8}\,Tr\int\,d^{4}xd^{2}\theta\,d^{2}\bar{\theta}\bigg(B\mathcal{D}^{2}V+\bar{B}\bar{\mathcal{D}}^{2}V-c_{\star}\mathcal{D}^{2}(sV)-\bar{c}_{\star}\bar{\mathcal{D}}^{2}(sV)\bigg)\,.
\label{21}
\end{eqnarray}
The super action above presents the chiral and antichiral pair of superfields $\{B$,$\bar{B}\}$ as Lagrange multipliers, where their classical equation of motions obey the constraint \eqref{SUPERFPYM1}. Moreover, the Faddeev-Popov ghost superfields' sector is settled by the pairs $\{c,c_{\star}\}$ representing the chiral ghost-antighost superfields  and $\{\bar{c},\bar{c}_{\star}\}$ the antichiral ones. After all these considerations, one has  
\begin{equation}
\mathcal{S}_{FP}=\mathcal{S}_{SYM}+\mathcal{S}_{gf}\,.
\label{superym}
\end{equation}
The action \eqref{superym} above obeys the following set of BRST transformations
\begin{eqnarray}
sV^{a} & = & (c^{a}-\bar{c}^{a})-\frac{i}{2}f^{abc}V^{b}(c^{c}+\bar{c}^{c})+\mathcal{O}(V^{2})\,,\nonumber\\
sc^{a} & = & \frac{i}{2}f^{abc}c^{b}c^{c}\,,\nonumber\\
sc^{a}_{\star} & = & B^{a}\,,\nonumber\\
sB^{a} & = & 0\,,\nonumber\\
s\bar{c}^{a} & = & \frac{i}{2}f^{abc}\bar{c}^{b}\bar{c}^{c}\,,\nonumber\\
s\bar{c}^{a}_{\star} & = & \bar{B}^{a}\,,\nonumber\\
s\bar{B}^{a} & = & 0\,.
\label{24}
\end{eqnarray}
From \eqref{24} above, one can observe the nonlinear BRST transformations of the set of superfields $(V^{a},c^{a},\bar{c}^{a})$. Since the superaction, eq. \eqref{superym}, is BRST-invariant and in order to deal with such nonlinear symmetry, we ought to introduce a set of external supersources, $(\Omega^{a},L^{a},\bar{L}^{a})$, by using the local composite operator formalism \cite{Piguet:1995er}, i.e.
\begin{eqnarray}
\mathcal{S}_{ext}[\Omega,L,\bar{L}] & = & Tr\bigg[-\frac{1}{8}\int\,d^{4}xd^{2}\theta\,d^{2}\bar{\theta}\bigg(\Omega(sV)\bigg)+\int d^{4}x\,d^{2}\theta\bigg(L(sc)\bigg)+\int d^{4}x\,d^{2}\bar{\theta}\bigg(\bar{L}(s\bar{c})\bigg)\bigg]\nonumber\\
\label{25}
\end{eqnarray}
where the aforementioned sources transform as BRST singlets
\begin{equation}
s\,\left(\,\Omega, L, \bar{L}\right)=  0\,.
\end{equation}
Finally, the BRST-invariant super Yang-Mills action assumes the following (complete) form
\begin{eqnarray}
\mathcal{S} & = & \mathcal{S}_{SYM} + \mathcal{S}_{gf} + \mathcal{S}_{ext}\nonumber\\
\mathcal{S} & = &  \frac{1}{128g^{2}}\,Tr\int\,d^{4}xd^{2}\theta\,\bigg(\bar{\mathcal{D}}^{2}\left(e^{-gV}\mathcal{D}_{\alpha}e^{gV}\right)\bar{\mathcal{D}}^{2}\left(e^{-gV}\mathcal{D}^{\alpha}e^{gV}\right)\bigg)\nonumber \\
& + & \frac{1}{128g^{2}}\,Tr\int\,d^{4}xd^{2}\bar{\theta}\,\bigg(\mathcal{D}^{2}\left(e^{-gV}\bar{\mathcal{D}}_{\dot{\alpha}}e^{gV}\right)\mathcal{D}^{2}\left(e^{-gV}\bar{\mathcal{D}}^{\dot{\alpha}}e^{gV}\right)\bigg)\nonumber \\
 & + & \frac{1}{8}\,Tr\int\,d^{4}xd^{2}\theta\,d^{2}\bar{\theta}\,\bigg(B\mathcal{D}^{2}V+\bar{B}\bar{\mathcal{D}}^{2}V-c_{\star}\mathcal{D}^{2}(sV)-\bar{c}_{\star}\bar{\mathcal{D}}^{2}(sV)\bigg)\nonumber \\
 & - & \frac{1}{8}\,Tr\int\,d^{4}xd^{2}\theta\,d^{2}\bar{\theta}\,\bigg(\Omega(sV)\bigg)+\,Tr\int d^{4}xd^{2}\theta\,\bigg(L\left(sc\right)\bigg)+\,Tr\int d^{4}xd^{2}\bar{\theta}\,\bigg(\bar{L}(s\bar{c})\bigg)\,.
 \label{27}
\end{eqnarray}
In the next subsection we will finally discuss about our proposal for the supersymmetric replica model. The careful reader will see that such sector has its own superfields and a BRST operator, where one will observe that with a convenient choice of parameters it will be possible writing an extended BRST operator that acts in all superfields from both sectors.

\subsection{The replica sector and an extended BRST operator}
As previously discussed, our model may be a good alternative
to shed some light on the study of some unexplained aspects of quantum confinement. Thus, as a first step to accomplish that we propose our SRM whose spirit of construction is analogous to the non-supersymmetric case reviewed in section \eqref{Rev}, i.e., we add a novel sector that mirrors eq.\eqref{27} with replica superfields. Indeed in order to realize that one may consider the following expression 
\begin{eqnarray}
\mathcal{S}_{SRM} & = & \frac{1}{128g^{2}}\,Tr\int\,d^{4}xd^{2}\theta\,\bigg(\bar{\mathcal{D}}^{2}\left(e^{-gV}\mathcal{D}_{\alpha}e^{gV}\right)\bar{\mathcal{D}}^{2}\left(e^{-gV}\mathcal{D}^{\alpha}e^{gV}\right)\bigg)\nonumber \\
& + & \frac{1}{128g^{2}}\,Tr\int\,d^{4}xd^{2}\bar{\theta}\,\bigg(\mathcal{D}^{2}\left(e^{-gV}\bar{\mathcal{D}}_{\dot{\alpha}}e^{gV}\right)\mathcal{D}^{2}\left(e^{-gV}\bar{\mathcal{D}}^{\dot{\alpha}}e^{gV}\right)\bigg)\nonumber \\
 & + & \frac{1}{8}\,Tr\int\,d^{4}xd^{2}\theta\,d^{2}\bar{\theta}\,\bigg(F\mathcal{D}^{2}\hat{V}+\bar{F}\bar{\mathcal{D}}^{2}\hat{V}-\hat{c}_{\star}\mathcal{D}^{2}(s\hat{V})-\hat{\bar{c}}_{\star}\bar{\mathcal{D}}^{2}(s\hat{V})\bigg)\nonumber \\
 & - & \frac{1}{8}\,Tr\int\,d^{4}xd^{2}\theta\,d^{2}\bar{\theta}\,\bigg(\hat{\Omega}(\omega\hat{V})\bigg)+\,Tr\int d^{4}x\,d^{2}\theta\,\bigg(\hat{L}\left(\omega\hat{c}\right)\bigg)+\,Tr\int d^{4}x\,d^{2}\bar{\theta}\,\bigg(\hat{\bar{L}}(\omega\hat{\bar{c}})\bigg)\,.\nonumber\\
 \label{28}
\end{eqnarray}
This replica super action is based on adding a completely analogous sector compared to eq.\eqref{27}. Therefore, eq.\eqref{28} also has its own vector superfield, $\hat{V}$, some Lagrange multipliers, $\{F$,$\bar{F}\}$; and the Faddeev-Popov ghost replica superfields $\{\hat{c},\hat{c}_{\star},\hat{\bar{c}},\hat{\bar{c}}_{\star}\}$. As a matter of fact, there is another BRST operator, namely $\omega$ responsible for acting only into the action \eqref{28}, i.e.
\begin{eqnarray}
\omega\hat{V}^{a} & = & (\hat{c}^{a}-\bar{c}^{a})-\frac{i}{2}f^{abc}\hat{V}^{b}(\hat{c}^{c}+\hat{\bar{c}}^{c})+\mathcal{O}(\hat{V}^{2})\,,\nonumber\\
\omega\hat{c}^{a} & = & \frac{i}{2}f^{abc}\hat{c}^{b}\hat{c}^{c}\,,\nonumber\\
\omega\hat{c}^{a}_{\star} & = & F^{a}\,,\nonumber\\
\omega F^{a} & = & 0\,,\nonumber\\
\omega\hat{\bar{c}}^{a} & = & \frac{i}{2}f^{abc}\hat{\bar{c}}^{b}\hat{\bar{c}}^{c}\,,\nonumber\\
\omega\hat{\bar{c}}^{a}_{\star} & = & \bar{F}^{a}\,,\nonumber\\
\omega\bar{F}^{a} & = & 0\,.
\label{128}
\end{eqnarray}

At this point, we could join both actions \eqref{27} and \eqref{28} into a single one, thus we may write it as \footnote{From now on we will consider $Tr(T^{a}T^{b})=k\delta^{ab}$ with $k= 1$.}
\begin{eqnarray}
\Sigma & = & \mathcal{S}+\mathcal{S}_{SRM}\nonumber\\
\Sigma & = & \frac{1}{128g^{2}}\,Tr\int\,d^{4}xd^{2}\theta\,\bigg(\bar{\mathcal{D}}^{2}\left(e^{-gV}\mathcal{D}_{\alpha}e^{gV}\right)\bar{\mathcal{D}}^{2}\left(e^{-gV}\mathcal{D}^{\alpha}e^{gV}\right)\bigg)\nonumber \\
& + & \frac{1}{128g^{2}}\,Tr\int\,d^{4}xd^{2}\bar{\theta}\,\bigg(\mathcal{D}^{2}\left(e^{-gV}\bar{\mathcal{D}}_{\dot{\alpha}}e^{gV}\right)\mathcal{D}^{2}\left(e^{-gV}\bar{\mathcal{D}}^{\dot{\alpha}}e^{gV}\right)\bigg)\nonumber \\
 & + & \frac{1}{8}Tr\int\,d^{4}xd^{2}\theta\,d^{2}\bar{\theta}\,\bigg(B\mathcal{D}^{2}V+\bar{B}\bar{\mathcal{D}}^{2}V-c_{\star}\mathcal{D}^{2}(sV)-\bar{c}_{\star}\bar{\mathcal{D}}^{2}(sV)\bigg)\nonumber \\
 & - & \frac{1}{8}Tr\int\,d^{4}xd^{2}\theta\,d^{2}\bar{\theta}\,\bigg(\Omega(sV)\bigg)+Tr\int d^{4}xd^{2}\theta\,\bigg(L\left(sc\right)\bigg)+Tr\int d^{4}xd^{2}\bar{\theta}\,\bigg(\bar{L}(s\bar{c})\bigg)\nonumber \\
 & + & \frac{1}{8}Tr\int\,d^{4}xd^{2}\theta\,d^{2}\bar{\theta}\,\bigg(F\mathcal{D}^{2}\hat{V}+\bar{F}\bar{\mathcal{D}}^{2}\hat{V}-\hat{c}_{\star}\mathcal{D}^{2}(\omega\hat{V})-\hat{\bar{c}}_{\star}\bar{\mathcal{D}}^{2}(\omega\hat{V})\bigg)\nonumber \\
 & - & \frac{1}{8}Tr\int\,d^{4}xd^{2}\theta\,d^{2}\bar{\theta}\,\bigg(\hat{\Omega}(\omega\hat{V})\bigg)+Tr\int d^{4}xd^{2}\theta\,\bigg(\hat{L}\left(\omega\hat{c}\right)\bigg)+Tr\int d^{4}xd^{2}\bar{\theta}\,\bigg(\hat{\bar{L}}(\omega\hat{\bar{c}})\bigg)\,.
 \label{29}
\end{eqnarray}
As written before, action $\eqref{29}$ above is invariant under the transformations of the two BRST operators,
\begin{equation}
    s\Sigma=\omega\Sigma=0\,.
\end{equation}
Hence, we may also get them together into a unique BRST extended operator described as
\begin{equation}
    \mathcal{Q}_{\epsilon}=s+\epsilon\,\omega\,,
\end{equation}
where one may see $\mathcal{Q}_{\epsilon}$ as a linear combination of the two BRST operators with $\epsilon$ as an arbitrary coefficient. All in all, the $\mathcal{Q}_{\epsilon}$ nilpotency is obtained from
\begin{equation}
    s^{2}=0\,,\qquad
    \omega^{2}=0\,,\qquad
    \{s,\omega\}=0\,.
\end{equation}
As the coefficient $\epsilon$ is completely arbitrary we will choose for simplicity $\epsilon=1$ and, as a consequence, our extended BRST will be given by
\begin{equation}
    \mathcal{Q}\equiv \mathcal{Q}_{\epsilon=1}
    =s+\omega\,.
\end{equation}
Then, the extended BRST transformations are defined in little sectors, namely
\begin{itemize}
\item{ The nonlinear BRST transformations:}
\begin{eqnarray}
\mathcal{Q}V^{a} & = &  (c^{a}-\bar{c}^{a})-\frac{i}{2}f^{acb}V^{b}(c^{c}+\bar{c}^{c})+\mathcal{O}(V^{2})\,,\nonumber \\
\mathcal{Q}c^{a} & = & -\frac{i}{2}f^{abc}c^{b}c^{c},\nonumber \\
\mathcal{Q}\bar{c}^{a} & = &-\frac{i}{2}f^{abc}\bar{c}^{b}\bar{c}^{c}\,,\nonumber \\
\mathcal{Q}\hat{V}^{a} & = &  (\hat{c}^{a}-\hat{\bar{c}}^{a})-\frac{i}{2}f^{acb}\hat{V}^{b}(\hat{c}^{c}+\hat{\bar{c}}^{c})+\mathcal{O}(\hat{V}^{2})\,,\nonumber \\
\mathcal{Q}\hat{c}^{a} & = & -\frac{i}{2}f^{abc}\hat{c}^{b}\hat{c}^{c}\,,\nonumber \\
\mathcal{Q}\hat{\bar{c}}^{a} & = & -\frac{i}{2}f^{abc}\hat{\bar{c}}^{b}\hat{\bar{c}}^{c}\,.
\end{eqnarray}
\item{ The BRST doublets:}
\begin{align}
\mathcal{Q}c_{\star}^{a}&=B^{a}\,,&
\mathcal{Q}\bar{c}_{\star}&=\bar{B}^{a}\,,&
\mathcal{Q}\hat{c}_{\star}^{a}&=F^{a}\,,&
\mathcal{Q}\hat{\bar{c}}_{\star}^{a}&=\bar{F}^{a}\,,\nonumber\\
\mathcal{Q}B^{a}&=0\,,&
\mathcal{Q}\bar{B}^{a}&=0\,,&
\mathcal{Q}F^{a}&= 0\,,&
\mathcal{Q}\bar{F}^{a}&=0\,.&
\end{align}
\item{The BRST singlets:}
\begin{equation}
\mathcal{Q}\,\left(\,\Omega^{a},
L^{a},
\bar{L}^{a},
\hat{\Omega}^{a},
\hat{L}^{a},
\hat{\bar{L}}^{a}\right) =  0\,.
\end{equation}
\end{itemize}
In addition, analogously with the non-supersymmetric case, the action eq. \eqref{29} is also invariant
according to the following (discrete) mirror symmetries,
\begin{eqnarray}
V^{a} & \rightarrow & \hat{V}^{a}|\hat{V}^{a}\rightarrow V^{a}\,,\nonumber \\
B^{a} & \rightarrow & F^{a}|F^{a}\rightarrow B^{a}\,,\nonumber \\
\bar{B}^{a} & \rightarrow & \bar{F}^{a}|\bar{F}^{a}\rightarrow\bar{B}^{a}\,,\nonumber \\
c^{a} & \rightarrow & \hat{c}^{a}|\hat{c}^{a}\rightarrow c^{a}\,,\nonumber\\
c_{\star}^{a} & \rightarrow & \hat{c}_{\star}^{a}|\hat{c}_{\star}^{a}\rightarrow c_{\star}^{a}\,,\nonumber\\ 
\bar{c}^{a} & \rightarrow & \hat{\bar{c}}^{a}|\hat{\bar{c}}^{a}\rightarrow\bar{c}^{a}\,,\nonumber 
\end{eqnarray}
\begin{eqnarray}
\bar{c}_{\star}^{a} & \rightarrow & \hat{\bar{c}}_{\star}^{a}|\hat{\bar{c}}_{\star}^{a}\rightarrow\bar{c}_{\star}^{a}\,,\nonumber \\
L^{a} & \rightarrow & \hat{L}^{a}|\hat{L}^{a}\rightarrow L^{a}\,,\nonumber \\
\bar{L}^{a} & \rightarrow & \hat{\bar{L}}^{a}|\hat{\bar{L}}^{a}\rightarrow\bar{L}^{a}\,,\nonumber \\
\Omega^{a} & \rightarrow & \hat{\Omega}^{a}|\hat{\Omega}^{a}\rightarrow\Omega^{a}\,.
\label{30}
\end{eqnarray}
Our aim at this moment is defining a connection between our model
with the super Gribov-Zwanziger formalism \cite{Amaral:2013uya} via tree level propagators. As one can infer from previous comments, in order to reproduce the i-particle propagators idea in the replica model, it is mandatory the
introduction of an interaction term responsible for coupling the vector superfield, $V$, with its replica, $\hat{V}$. Once ready, the tree-level propagators will change in such a way that they violate Källén-Lehnmman representation for physical excitations. Thus, we defined the terms in charge of mixing both SYM and SRM sectors as 
\begin{eqnarray}
\Sigma_{mixt} & = & \frac{1}{8}\int\,d^{4}xd^{2}\theta\,d^{2}\bar{\theta}\bigg(m^{2}\delta^{ab}\left(V^{a}V^{b}+\hat{V}^{a}\hat{V}^{b}\right)+i\mu^{2}\delta^{ab}V^{a}\hat{V}^{b}\bigg)\,.
\label{32}
\end{eqnarray}
As the BRST symmetry is softly broken when one mixes such superfields, we shall complete our action \eqref{29} by adding two more doublets of external supersources \footnote{This procedure is done since sometime in the near future we would like to prove that such model is renormalizable at all orders of a loop expansion.}. Then, one has
\begin{align}
\mathcal{Q}R^{ab}&=J^{ab}\,,&
\mathcal{Q}\mathcal{N}^{ab}&=\mathcal{M}^{ab}\,,\nonumber\\
\mathcal{Q}J^{ab}&=0\,,&
\mathcal{Q}\mathcal{M}^{ab}&=0\,.&
\end{align}
In a convenient way, one may set all these sources at some specific values known as the physical limit of the model, therefore
\begin{eqnarray}
R_{|Phys}^{ab} & = & 0\,,\nonumber\\
J_{|Phys}^{ab} & = & m^{2}\delta^{ab}\,,\nonumber\\
\mathcal{N}_{|Phys}^{ab} & = & 0\,,\nonumber\\
\mathcal{M}_{|Phys}^{ab} & = & i\mu^{2}\delta^{ab}\,.
\label{33}
\end{eqnarray}
Thus, one can describe the action \eqref{32} in a BRST-like type as
\begin{eqnarray}
\Sigma_{mixt} & = & \frac{1}{8}\int\,d^{4}xd^{2}\theta\,d^{2}\bar{\theta}\bigg(R^{ab}\mathcal{Q}(V^{a}V^{b})+J^{ab}(V^{a}V^{b}+\hat{V}^{a}\hat{V}^{b})+\mathcal{M}^{ab}V^{a}\hat{V}^{b}+\mathcal{N}^{ab}\mathcal{Q}(V^{a}V^{b})\bigg)\,.\nonumber\\
\end{eqnarray}
Also, to make things easier to the curious reader, one may write such terms only in function of the space-time components, for instance 
\begin{eqnarray}
\frac{1}{8}\int\,d^{4}xd^{2}\theta\,d^{2}\bar{\theta}\,\bigg(\,J^{ab}V^{a}V^{b}\bigg) & = & \frac{1}{8}\int d^{4}x\,\mathcal{D}^{2}\bar{\mathcal{D}}^{2}\bigg(\,J^{ab}V^{a}V^{b}\bigg)\propto\int d^{4}x\,\bigg(\,A_{\mu}^{a}A_{\mu}^{b}\bigg)\,,\nonumber \\
\frac{1}{8}\int\,d^{4}xd^{2}\theta\,d^{2}\bar{\theta}\,\bigg(\,J^{ab}\hat{V}^{a}\hat{V}^{b}\bigg) & = & \frac{1}{8}\int d^{4}x\,\mathcal{D}^{2}\bar{\mathcal{D}}^{2}\bigg(\,J^{ab}\hat{V}^{a}\hat{V}^{b}\bigg)\propto\int d^{4}x\,\bigg(\,\hat{A}_{\mu}^{a}\hat{A}_{\mu}^{b}\bigg)\,,\nonumber \\
\frac{1}{8}\int\,d^{4}xd^{2}\theta\,d^{2}\bar{\theta}\,\bigg(\,\mathcal{M}^{ab}V^{a}\hat{V}^{b}\bigg) & = & \frac{1}{8}\int d^{4}x\,\mathcal{D}^{2}\bar{\mathcal{D}}^{2}\bigg(\,\mathcal{M}^{ab}V^{a}\hat{V}^{b}\bigg)\propto\int d^{4}x\,\bigg(\,A_{\mu}^{a}\hat{A}_{\mu}^{b}\bigg)\,.\nonumber \\
\end{eqnarray}
Once we have applied all those steps, we can write down our complete super action which encloses all the necessary superfields, supersources and their corresponding super replicas in the following way
\begin{eqnarray}
\Sigma_{full} & = & \frac{1}{128g^{2}}\,Tr\int\,d^{4}xd^{2}\theta\,\bigg(\bar{\mathcal{D}}^{2}\left(e^{-gV}\mathcal{D}_{\alpha}e^{gV}\right)\bar{\mathcal{D}}^{2}\left(e^{-gV}\mathcal{D}^{\alpha}e^{gV}\right)\bigg)\nonumber \\
& + & \frac{1}{128g^{2}}\,Tr\int\,d^{4}xd^{2}\bar{\theta}\,\bigg(\mathcal{D}^{2}\left(e^{-gV}\bar{\mathcal{D}}_{\dot{\alpha}}e^{gV}\right)\mathcal{D}^{2}\left(e^{-gV}\bar{\mathcal{D}}^{\dot{\alpha}}e^{gV}\right)\bigg)\nonumber \\
 & + & \frac{1}{8}\,Tr\int\,d^{4}xd^{2}\theta\,d^{2}\bar{\theta}\,\bigg(B\mathcal{D}^{2}V+\bar{B}\bar{\mathcal{D}}^{2}V-c_{\star}\mathcal{D}^{2}(\mathcal{Q}V)-\bar{c}_{\star}\bar{\mathcal{D}}^{2}(\mathcal{Q}V)\bigg)\nonumber \\
 & - & \frac{1}{8}Tr\int\,d^{4}xd^{2}\theta\,d^{2}\bar{\theta}\,\bigg(\Omega(\mathcal{Q}V)\bigg)+Tr\int d^{4}xd^{2}\theta\,\bigg(L\left(\mathcal{Q}c\right)\bigg)+Tr\int d^{4}xd^{2}\bar{\theta}\,\bigg(\bar{L}(\mathcal{Q}\bar{c})\bigg)\nonumber \\
 & + & \frac{1}{8}Tr\int\,d^{4}xd^{2}\theta\,d^{2}\bar{\theta}\,\bigg(F\mathcal{D}^{2}\hat{V}+\bar{F}\bar{\mathcal{D}}^{2}\hat{V}-\hat{c}_{\star}\mathcal{D}^{2}(\mathcal{Q}\hat{V})-\hat{\bar{c}}_{\star}\bar{\mathcal{D}}^{2}(\mathcal{Q}\hat{V})\bigg)\nonumber \\
 & - & \frac{1}{8}Tr\int\,d^{4}xd^{2}\theta\,d^{2}\bar{\theta}\,\bigg(\hat{\Omega}(\mathcal{Q}\hat{V})\bigg)+Tr\int d^{4}xd^{2}\theta\,\bigg(\hat{L}\left(\mathcal{Q}\hat{c}\right)\bigg)+Tr\int d^{4}xd^{2}\bar{\theta}\,\bigg(\hat{\bar{L}}(\mathcal{Q}\hat{\bar{c}})\bigg)\nonumber \\
 & + & \frac{1}{8}Tr\int\,d^{4}xd^{2}\theta\,d^{2}\bar{\theta}\,\bigg(R\mathcal{Q}(VV)+J(VV+\hat{V}\hat{V})+\mathcal{M}V\hat{V}+\mathcal{N}\mathcal{Q}(VV)\bigg)\,.\nonumber \\
\label{36}
\end{eqnarray}
With eq.\eqref{36} addressed above we could study the algebraic renormalization procedure \cite{Piguet:1995er} for our model, however this work will be done sometime in the near future \cite{Toledo}. In the present article we will limit ourselves to work only on the quadratic part of eq.\eqref{36} to be shown in the following section.

\section{The gluon and gluino tree-level propagators}
\label{prop}

Here, as previously mentioned, we will compute the gluon and gluino tree-level propagators and compare them with the ones obtained from the Gribov-Zwanziger supersymmetric extension. Also, such procedure will enable us finding the same results for space-time action when one compares with the original non-supersymmetric replica model. Then, in order to do that, we ought to present as a first step the quadratic part of the action \eqref{36} below,
\begin{eqnarray}
\Sigma_{quad} & = & \frac{1}{128g^{2}}\int d^{4}x\,d^{2}\theta\,\bigg(V^{a}\mathcal{D}^{\alpha}\bar{\mathcal{D}}^{2}\mathcal{D}_{\alpha}V^{a}+\hat{V}^{a}\mathcal{D}^{\alpha}\bar{\mathcal{D}}^{2}\mathcal{D}_{\alpha}\hat{V}^{a}\bigg)\nonumber \\
 & + & 
 \frac{1}{128g^{2}}\int d^{4}x\,d^{2}\bar{\theta}\,\bigg(
 V^{a}\bar{\mathcal{D}}_{\dot{\alpha}}\mathcal{D}^{2}\bar{\mathcal{D}}^{\dot{\alpha}}V^{a}+\hat{V}^{a}\bar{\mathcal{D}}_{\dot{\alpha}}\mathcal{D}^{2}\bar{\mathcal{D}}^{\dot{\alpha}}\hat{V}^{a}\bigg)\nonumber \\
 & + & \frac{1}{8}\int\,d^{4}xd^{2}\theta\,d^{2}\bar{\theta}\,\bigg[B^{a}\mathcal{D}^{2}V^{a}+\bar{B}^{a}\bar{\mathcal{D}}^{2}V^{a}+F^{a}\mathcal{D}^{2}\hat{V}^{a}+\bar{F}^{a}\bar{\mathcal{D}}^{2}\hat{V}^{a}\nonumber \\
 & + & m^{2}\bigg(V^{a}V^{a}+\hat{V}^{a}\hat{V}^{a}\bigg)+i\mathcal{\mu}^{2}V^{a}\hat{V}^{a}\bigg]\,.
 \label{45}
\end{eqnarray}
Two important remarks are needed to be done at this point: the first one is that in the case where the values of $m$ and $\mu$ become negligible when one compares to the energy scale, the two sectors of the action \eqref{36} decouple, giving rise a two independent models (such event also occurs in the non-supersymmetric case). Moreover, one can see from eq. \eqref{45} that the ghosts terms don't appear since they are decoupled terms, thus the action is written without their presence and we can focus only on the super-gauge field sector. As a second step we describe the generator functional for all the correlations functions as
\begin{eqnarray}
Z[\mathcal{J}] & = & \int\,D\Phi\exp\bigg(-\Sigma_{quad}-\mathcal{J}\Phi\bigg)\,,
\end{eqnarray}
where $\Phi= (V,\hat{V},B,F,\bar{B},\bar{F})$ and $\mathcal{J}=(J_{V},J_{\hat{V}},J_{B},J_{F},J_{\bar{B}},J_{\bar{F}})$ represent all the superfields and Schwinger supersources of the action eq.\eqref{45}, respectively. The generating functional for connected diagrams is given by the logarithm
of Z, i.e.
\begin{eqnarray}
W[\mathcal{J}] & = & \ln\,Z[\mathcal{J}]\,.
\end{eqnarray}
Hereby, $W$ may be viewed as the collection of
all Green's functions when the Taylor series expansions with respect
to external sources are taken.
For this work it is not important but we can express $W$ in terms of the generator functional $\Gamma$ responsible for describing all one particle irreducible ($1$PI) diagrams,
\begin{eqnarray}
\Gamma\left[\Phi\right]& = & W\left[\mathcal{{J}}\right] -\int\,d^{4}x\,d^{2}\theta\,d^{2}\bar{\theta}\,\bigg(J_{V}V+J_{\hat{V}}\hat{V}\bigg)-\int\,d^{4}x\,d^{2}\theta\,\bigg(J_{B}B+J_{F}F\bigg)\nonumber\\
&-&\int\,d^{4}x\,d^{2}\bar{\theta}\,\bigg(J_{\bar{B}}\bar{B}+J_{\bar{F}}\bar{F}\bigg)\,.
\label{48}
\end{eqnarray}
From now on, we will use the connected generation functional $W$ in order to fulfill our goal of computing the gluon and gluino tree-level propagators. Therefore, we have the following relations
\begin{eqnarray}
V=\frac{\delta W}{\delta J_{V}}\,, & \quad\quad\quad\quad\,\, & \hat{V}=\frac{\delta W}{\delta J_{\hat{V}}}\,,\nonumber \\
B=\frac{\delta W}{\delta J_{B}}\,, & \quad\quad\quad\quad\,\, & F=\frac{\delta W}{\delta J_{F}}\,,\nonumber \\
\bar{B}=\frac{\delta W}{\delta J_{\bar{B}}}\,, & \quad\quad\quad\quad\,\, & \bar{F}=\frac{\delta W}{\delta J_{\bar{F}}}\,.
\end{eqnarray}
At this moment, we will adopt here the same strategy developed by \cite{Piguet:1996ys} for SYM action, in this way the propagator is defined as
\begin{eqnarray}
\frac{\delta^{2}W}{\delta J_{V}\delta J_{V}}_{|J_{V}=0} & = & -\frac{1}{i}\langle VV\rangle\,.
\end{eqnarray}
Now, we will present the equations of motion as another step in order to compute the propagators, i.e.
\begin{itemize}
    \item {The equation of motion for the supergauge field $V$:}
\begin{eqnarray}
-J_{V}^{a} & = & \frac{1}{64g^{2}}\mathcal{D}^{\alpha}\bar{\mathcal{D}}^{2}\mathcal{D}_{\alpha}V^{a}+\frac{1}{8}\mathcal{D}^{2}B^{a}+\frac{1}{8}\bar{\mathcal{D}}^{2}\bar{B}^{a}+\frac{1}{8}m^{2}V^{a}+\frac{1}{8}i\mathcal{\mu}^{2}\hat{V}^{a}\,.
\end{eqnarray}
\item {The equation of motion for the replica supergauge field $\hat{V}$:}
\begin{eqnarray}
-J_{\hat{V}}^{a} & = & \frac{1}{64g^{2}}\mathcal{D}^{\alpha}\bar{\mathcal{D}}^{2}\mathcal{D}_{\alpha}\hat{V}^{a}+\frac{1}{8}\mathcal{D}^{2}\hat{B}^{a}+\frac{1}{8}\bar{\mathcal{D}}^{2}\hat{B}^{a}+\frac{1}{8}m^{2}\hat{V}^{a}+\frac{1}{8}i\mathcal{\mu}^{2}V^{a}\,.
\end{eqnarray}
\item {The equations of motion for the chiral and antichiral auxiliary superfields $(B,\bar{B})$:}
\begin{eqnarray}
-J_{B}^{a} =  \frac{1}{8}\bar{\mathcal{D}}^{2}\mathcal{D}^{2}V^{a}\,,& \quad\quad\quad\quad\,\, & -J_{\bar{B}}^{a}  =  \frac{1}{8}\mathcal{D}^{2}\bar{\mathcal{D}}^{2}V^{a}\,.
\end{eqnarray}
\item {The equations of motion for the chiral and antichiral auxiliary replica superfields $(F,\bar{F})$:}
\begin{eqnarray}
-J_{F}^{a} =  \bar{\mathcal{D}}^{2}\mathcal{D}^{2}\hat{V}^{a}\,,& \quad\quad\quad\quad\,\, & -J_{\bar{F}}^{a} =  \mathcal{D}^{2}\bar{\mathcal{D}}^{2}\hat{V}^{a}\,.
\end{eqnarray}
\end{itemize}
As we are interested in the gluon and gluino tree-level propagators, we ought to first focus only on $\langle VV\rangle$, $\langle V\hat{V}\rangle$ and $\langle \hat{V}\hat{V}\rangle$ results. Then, after few steps we have
\begin{eqnarray}
\langle V^{a}V^{b}\rangle & = & \frac{8g^{2}i\left(\partial^{2}+g^{2}m^{2}\right)P^{T}\delta^{ab}\delta_{V}}{\partial^{4}+2g^{2}m^{2}\partial^{2}+g^{4}\left(m^{4}+\mu^{4}\right)}\,.
\label{P1}
\end{eqnarray}
From the well-known now mirror symmetry one has the same expression presented in \eqref{P1} for the replica vector superfield propagator, i.e.
\begin{eqnarray}
\langle{\hat{V}^{a}\hat{V}^{b}}\rangle & = & \frac{8g^{2}i\left(\partial^{2}+g^{2}m^{2}\right)P^{T}\delta^{ab}\delta_{V}}{\partial^{4}+2g^{2}m^{2}\partial^{2}+g^{4}\left(m^{4}+\mu^{4}\right)}\,,
\label{P2}
\end{eqnarray}
and the propagator which encodes the mixing term of both sectors is written as
\begin{eqnarray}
\langle V^{a}\hat{V}^{b}\rangle & =-\frac{8ig^{4}\mathcal{\mu}^{2}}{\left[\partial^{2}+g^{2}m^{2}\right]} & \bigg(\frac{\left(\partial^{2}+g^{2}m^{2}\right)P^{T}\delta^{ab}\delta_{V}}{\partial^{4}+2g^{2}m^{2}\partial^{2}+g^{4}\left(m^{4}+\mu^{4}\right)}\bigg)\,,
\label{P3}
\end{eqnarray}
where $P^{T}$ and $\delta_{V}$ are given in appendix \eqref{conv}. From eqs. \eqref{P1}, \eqref{P2} and \eqref{P3} above, one can note that the propagators have an analogous structure as the Gribov-Zwanziger case in space components
giving rise to the earlier mentioned positivity violation result of Källén-Lehnman's spectral representation. Such event is
observed in the Gribov-Zwanziger supersymmetric extension \cite{Amaral:2013uya} when the vector superfield is decomposed in its components. Indeed, as the reader can imagine, in order to obtain a confined model-type we will follow the same procedure in this article, then we shall select the fermionic coordinates $(\theta,\bar{\theta})$ terms related to both gluon and gluino's sectors, i.e. their local composite operators components will be obtained by using the transverse operator $P^{T}$, thus one has
\begin{eqnarray}
\theta^{\delta}\sigma_{\delta\dot{\delta}}^{\mu}\bar{\theta}^{\dot{\delta}}\theta^{\eta}\sigma_{\eta\dot{\eta}}^{\nu}\bar{\theta}^{\dot{\eta}} & \longrightarrow & AA,\quad\hat{A}\hat{A},\quad A\hat{A}\,,\\
\bar{\theta}^{2}\theta^{\alpha}\theta^{2}\bar{\theta}_{\dot{\alpha}} & \longrightarrow & \lambda\bar{\lambda},\quad\hat{\lambda}\hat{\bar{\lambda}},\quad\lambda\hat{\bar{\lambda}},\quad\bar{\lambda}\hat{\lambda}\,.
\end{eqnarray}
For instance, let us work on the result \eqref{P1}, where
we ought to identify the gluon propagator and of course the gluino one, to accomplish such aim we have too make explicit all components of the transverse projector whose description in terms of the fermionic  coordinates is given by
\begin{eqnarray}
P^{T} & = & \frac{1}{8\partial^{2}}\left(1+\frac{1}{4}\left(\theta_{1}-\theta_{2}\right)^{2}\left(\bar{\theta}_{1}-\bar{\theta}_{2}\right)^{2}\partial^{2}\right)\left(1+\left(\theta_{1}\sigma\bar{\theta}_{2}-\theta_{2}\sigma\bar{\theta}_{1}\right)\partial_{\mu}+\left(\theta_{1}\sigma\bar{\theta}_{2}-\theta_{2}\sigma\bar{\theta}_{1}\right)^{2}\partial^{2}+...\right)\nonumber\\
&\times&\delta^{4}\left(x_{1}-x_{2}\right)\,,
\end{eqnarray}
and apply it, as mentioned before, on eq.\eqref{P1}, thus we can finally figure out the gluon and gluino propagators expressions written in space components as follows
\begin{itemize}
    \item Gluon:
\begin{eqnarray}
\langle A_{\mu}^{a}A_{\nu}^{b}\rangle & = & \frac{g^{2}i\left[\partial^{2}+g^{2}m^{2}\right]\delta^{ab}\delta_{V}}{4\Big(\partial^{4}+2g^{2}m^{2}\partial^{2}+g^{4}\left(m^{4}+\mu^{4}\right)\Big)}\left(\delta_{\mu\nu}-2\frac{\partial_{\nu}\partial_{\mu}}{\partial^{2}}\right)\,.
\label{gluon}
\end{eqnarray}
\item Gluino:
\begin{eqnarray}
\langle \lambda^{a\,\alpha}\bar{\lambda}^{b}_{\alpha}\rangle & = & \frac{5g^{2}i\left[\partial^{2}+g^{2}m^{2}\right]\sigma^{\mu}\partial_{\mu}\delta^{ab}\delta_{V}}{2\Big(\partial^{4}+2g^{2}m^{2}\partial^{2}+g^{4}\left(m^{4}+\mu^{4}\right)\Big)}\,.
\label{sgluon}
\end{eqnarray}
\end{itemize}
From the results above, namely \eqref{gluon} and \eqref{sgluon}, one can note the confining behavior either for gluon or for gluino propagators quite analogous to the super GZ model presented in \cite{Amaral:2013uya} as expected since the beginning of this work. 

\section{Conclusions}
\label{concl}
Initially, we have reviewed the non-supersymmetric replica model and its propagators with a confining character. Then, we have proposed a $\mathcal{N}=1$ supersymmetric extension of the replica model which may be considered as a novel and alternative approach in order to study the confinement phenomenon of the elementary degrees of freedom (i.e. gluons and gluinos). Also, our proposal was based on doubling the super Yang-Mills action quantized in the Landau gauge, i.e. we have added mirror terms in the original action, as a consequence, we had two BRST operators, one for each sector, that could be joined into a single one by a convenient choice of parameters giving rise to an extended BRST symmetry. In fact, such construction was fundamental to introduce the massive terms at physical limit, namely $(m^{2},i\mu^{2})$, which were responsible for connecting both sectors in a BRST-invariant way when replaced by a certain set of external sources, whereas the invariance of this (super)symmetry is a basic condition to know whether our model is renormalizable to all orders in perturbation theory, a subject that is currently under investigation \cite{Toledo}.

In the second part, since the gluons and gluinos propagators' structure are analogous to the results computed for super Gribov-Zwanziger model \cite{Amaral:2013uya}, we could observe that there is an equivalence between both frameworks (an event that also occurs for the non-supersymmetric case). Thus, we can consider our massive parameters $(m^{2},i\mu^{2})$ equivalents to the well-known Gribov parameter, hence we were able to access the infrared region by generating the confining-type propagators $\langle AA \rangle$ and $\langle \bar{\lambda}\lambda \rangle$, as shown by the equations \eqref{gluon} and \eqref{sgluon}, of course such results fulfill our aim and ensured that we have constructed an alternative model with a distinct viewpoint involving the gluons and gluinos and it could be considered in principle as a confined-type way. However, the discussion about confining models and SUSY breaks still lacks a better understanding.  

Finally, we point out some potential investigations related to our SRM. First, as already written before we will use the algebraic renormalization procedure in order to prove whether our model is renormalizable or not. Moreover, we would like to extend such proposal for $\mathcal{N}=2$ and $\mathcal{N}=4$, where the first case can be considered an interesting study due to possible comparisons with Seiberg and Witten model \cite{Seiberg:1994rs} as well as for $\mathcal{N}=4$ where we could investigate potential novel non-perturbative features and compare them with Maldacena's conjecture \cite{Maldacena:1997re}. Also, we consider this work as a first step to shed some light on the fermions introduction in the non-supersymmetric replica model. Any progress in those directions will be reported sometime in the near future.

\section*{Acknowledgments}
This study was financed by the National Council for Scientific and Technological Development (CNPq/MCTI). R. C. Terin is supported through the Junior Postdoctoral Fellowship Program (PDJ/CNPq), Finance Code - 151397/2020-1; M. A. L. Capri is a level PQ-2 researcher under
the program Produtividade em Pesquisa (CNPq), Finance Code - 313068/2020-8. H. C. Toledo would like to thank the Universidade do Estado do Rio de Janeiro (UERJ) for its hospitality during his PhD.



\appendix
\section{Notations and conventions}
\label{conv}
In this appendix we will show our notations and conventions that we use in order to propose our SRM and compute the gluon and gluino propagators at tree-level. In addition, such rules used here are the same as those described in \cite{Piguet:1996ys}. For instance, in Euclidean space the metric is defined as
\begin{equation}
\delta_{\mu\nu} = diag(+,+,+,+)\,.
\end{equation}
From SUSY's literature one can observe the ordinary derivative acting in a superfield does not assume the characteristic of a superfield, thus one needs to redefine it with the aim of preserving the superfield's structure, i.e.
\begin{eqnarray}
	\mathcal{D}_{\alpha} & = & \frac{\partial}{\partial\theta^{\alpha}}+i\sigma_{\alpha\dot{\alpha}}^{\mu}\bar{\theta}^{\dot{\alpha}}\partial_{\mu}\,,\\
	\bar{\mathcal{D}}_{\dot{\alpha}} & = & -\frac{\partial}{\partial\bar{\theta}^{\dot{\alpha}}}-i\theta^{\alpha}\sigma_{\alpha\dot{\alpha}}^{\mu}\partial_{\mu}\,.
\end{eqnarray}
These operators have some properties described below
\begin{eqnarray}
\left\{\mathcal{D}_{\alpha},\bar{\mathcal{D}}_{\dot{\alpha}}\right\}  & = & -2i\sigma_{\alpha\dot{\alpha}}^{\mu}\partial_{\mu}\,,\\
\left[\mathcal{D}_{\alpha},\bar{\mathcal{D}}_{\dot{\alpha}}\right] 
& = & -8i\left(\bar{\mathcal{D}}_{\dot{\alpha}}\sigma^{\alpha\dot{\alpha}\mu}\mathcal{D}_{\alpha}\right)\partial_{\mu}+16\partial^{2}\,,
\end{eqnarray}
where
\begin{eqnarray}
\bar{\sigma}^{\mu\alpha\dot{\alpha}} & = & \epsilon^{\dot{\alpha}\dot{\beta}}\epsilon^{\alpha\beta}\sigma_{\beta\dot{\beta}}^{\mu}\,,\\
\sigma_{\dot{\alpha}\alpha}^{\mu}\bar{\sigma}^{\mu\dot{\beta}\beta} & = & -2\delta_{\alpha}^{\beta}\delta_{\dot{\alpha}}^{\dot{\beta}}\,.
\end{eqnarray}
Moreover, the Grassmann coordinates obey the following algebra
\begin{eqnarray}
\theta^{\alpha}\theta^{\beta} &=& -\frac{1}{2}\epsilon^{\alpha \beta}\theta^{2}\,,\\
\bar{\theta}^{\dot{\alpha}}\bar{\theta}^{\dot{\beta}} &=& -\frac{1}{2}\epsilon^{\dot{\alpha} \dot{\beta}}\theta^{2}\,.
\end{eqnarray}
Another useful property about the covariant derivatives is its connection with the integral over the Grassmann coordinates, i.e.
\begin{eqnarray}
\int\,d^{4}xd^{2}\theta\,d^{2}\bar{\theta} = \int d^{4}x d^{2}\bar{\theta} d^{2}{\theta} = \int d^{4}x\,\mathcal{D}^{2}\bar{\mathcal{D}}^{2} = \int d^{4}x \bar{\mathcal{D}}^{2}\mathcal{D}^{2}\,.
\end{eqnarray}
Finally, we have below all the projection operators that helped us  compute the gluon and gluino propagators
\begin{eqnarray}
P^{T} & = & \frac{\mathcal{D}^{\alpha}\bar{\mathcal{D}}^{2}\mathcal{D}_{\alpha}}{8\partial^{2}}=\frac{\bar{\mathcal{D}}_{\dot{\alpha}}\mathcal{D}^{2}\bar{\mathcal{D}}^{\dot{\alpha}}}{8\partial^{2}}\,,\nonumber \\
P^{L} & = & P_{+}^{L}+P_{-}^{L}=-\frac{\mathcal{D}^{2}\bar{\mathcal{D}}^{2}+\bar{\mathcal{D}}^{2}\mathcal{D}^{2}}{16\partial^{2}}\,,\nonumber \\
P_{+}^{L} & = & -\frac{\mathcal{D}^{2}\bar{\mathcal{D}}^{2}}{16\partial^{2}}\,,\nonumber \\
P_{-}^{L} & = & -\frac{\bar{\mathcal{D}}^{2}\mathcal{D}^{2}}{16\partial^{2}}\,,\nonumber \\
\mathcal{D}^{2}\bar{\mathcal{D}}^{2}\mathcal{D}^{2} & = & -16\partial^{2}\mathcal{D}^{2}\,,\nonumber\\
\bar{\mathcal{D}}^{2}\mathcal{D}^{2}\bar{\mathcal{D}}^{2} & = & -16\partial^{2}\bar{\mathcal{D}}^{2}\,,\nonumber \\
P^{T}P^{L} & = & 0\,,\nonumber \\
(P^{T})^{2} & = & P^{T}\,,\nonumber 
\end{eqnarray}
\begin{eqnarray}
(P^{L})^{2} & = & P^{L}\,,\nonumber \\
P^{T}+P^{L} & = & 1\,,\nonumber\\
\mathcal{D}^{\alpha}\bar{\mathcal{D}}^{2}\mathcal{D}_{\alpha} & = & \bar{\mathcal{D}}_{\dot{\alpha}}\mathcal{D}^{2}\bar{\mathcal{D}}^{\dot{\alpha}}\,,\nonumber \\
\delta_{V} & = & \delta^{4}(x-y)\delta^{2}(\theta_{1}-\theta_{2})\delta^{2}(\bar{\theta}_{1}-\bar{\theta}_{2})\,,\nonumber \\
\delta_{S} & = & \delta^{4}(x-y)\delta^{2}(\theta_{1}-\theta_{2})\nonumber\,,\\
\delta_{\bar{S}} & = & \delta^{4}(x-y)\delta^{2}(\bar{\theta}_{1}-\bar{\theta}_{2})\,. \label{propriedades}
\end{eqnarray}




\end{document}